\documentclass[aps,preprintnumbers,prl,twocolumn,superscriptaddress]{revtex4}

\usepackage{epsfig,latexsym,cancel,amssymb,amsmath}
\usepackage{graphicx}
\usepackage{epstopdf}

\DeclareMathOperator{\real}{Re}
\DeclareMathOperator{\imag}{Im}

\newcommand{\eV}{\ensuremath{\mathrm{eV}}}

\newcommand{\SMgroup}{\ensuremath {\mathrm{SU}}(3)_{c}\times {\mathrm{SU}}(2)_{L}\times{\mathrm U}(1)_{Y}}
\newcommand{\Uoneprime}{\ensuremath {\mathrm{U}(1)_V}}

\begin{document}

\preprint{PI-PARTPHYS-324}

\title{Dark Matter Detectors as Dark Photon Helioscopes}

 \author{Haipeng An}
\affiliation{Perimeter Institute, Waterloo, Ontario N2L 2Y5, Canada}

\author{Maxim Pospelov}
\affiliation{Perimeter Institute, Waterloo, Ontario N2L 2Y5, Canada}
\affiliation{Department of Physics and Astronomy, University of Victoria, Victoria, BC, V8P 5C2, Canada}

\author{Josef Pradler}
\affiliation{Department of Physics and Astronomy, Johns Hopkins University, Baltimore, MD 21210, USA}


\begin{abstract}
  Light new particles with masses below 10 keV, often considered as a
  plausible extension of the Standard Model, will be emitted from the
  solar interior, and can be detected on the Earth with a variety of
  experimental tools.  Here we analyze the new "dark" vector state
  $V$, a massive vector boson mixed with the photon via an angle
  $\kappa$, that in the limit of the small mass $m_V$ has its emission
  spectrum strongly peaked at low energies.  Thus, we utilize the
  constraints on the atomic ionization rate imposed by the results of
  the XENON10 experiment to set the limit on the parameters of this
  model: $\kappa \times m_V< 3\times10^{-12}\,\eV$.  This makes low-threshold
  Dark Matter experiments the most sensitive dark vector helioscopes,
  as our result not only improves current experimental bounds from
  other searches by several orders of magnitude, but also surpasses even
  the most stringent astrophysical and cosmological limits in a
  seven-decade-wide interval of $m_V$.  We generalize this approach to
  other light exotic particles, and set the most stringent direct
  constraints on "mini-charged" particles.
\end{abstract}

\maketitle

\paragraph{Introduction}
\label{sec:introduction}

The Standard Model (SM) of particle physics based on the gauge group
structure $G_{SM}=\SMgroup$ and the Higgs mechanism is now firmly
established and confirmed in a wide range of energies.  At the same
time there are reasons to think that SM is an effective theory, and
new ingredients must be added to it.  New states may exist both at
higher energy scales with sizable couplings to SM, and at low energies
where such states would have to be neutral under $G_{SM}$ and very
weakly coupled to the SM particles. 
Among the few distinct classes to couple new light states to the SM
singlet operators, the U(1)$_Y$ hypercharge field strength appears as
the most natural~\cite{Holdom:1985ag}. It is singled out not only by
its minimality but by its enhancement in the infrared~(IR). The
hypercharge portal leads to the mixing of an additional U(1)$_{V}$
gauge boson (called "dark photon" from here on) with the SM photon,
and thus can easily manifest itself in low-energy phenomena.

In the last few years, the model of kinetically mixed vectors has
received tremendous attention, theoretically as well as
experimentally. While the mass range above $\sim1$ MeV is mostly
subjected to traditional particle physics constraints with
high-intensity beams, the intermediate mass range, 10~eV to 1~MeV, is
much constrained by astrophysics and cosmology.  In the lowest mass
range, $m_V < 10 $ eV, astrophysical limits are complemented by direct
laboratory searches of dark photons in non-accelerator type
experiments.  A collection of low-energy constraints on dark photons
can be found in the recent review~\cite{Jaeckel:2010ni}. Among the
most notable detection strategies are the
``light-shining-through-wall'' experiments (LSW)~\cite{Ahlers:2007qf}
and the conversion experiments from the solar dark photon flux,
``helioscopes''~\cite{Redondo:2008aa}. The latter class of experiments
derives its sensitivity from the fact that such light vectors are
readily excited in astrophysical environments, such as
\textit{e.g.}~in the solar interior, covering a wide range of masses
up to $m_V \sim $ few keV.  Stellar astrophysics provides stringent
constraints on any type of light, weakly-interacting particles when
the emission becomes kinematically
possible~\cite{Raffelt:1996wa}. Only in a handful of examples does the
sensitivity of terrestrial experiments match the stellar energy loss
constraints.

In a recent work~\cite{An:2013yfc} we have identified a new stellar
energy loss mechanism originating from the resonant production of
longitudinally polarized dark photons.  Ref.~\cite{An:2013yfc}
significantly improved limits on dark photons compared to the original
analysis~\cite{Redondo:2008aa}, to the extent that all current LSW and
helioscope experiments now find themselves deep inside astrophysically
excluded regions.

The purpose of this letter is to show that the newly calculated flux
of dark photons in combination with utmost sensitivity of direct Dark
Matter detection experiments to atomic ionization make a powerful
probe of dark photon models.  In what follows, we calculate the solar
flux of dark photons, both for the case of a ``hard'' Stueckelberg
mass $m_V$ and for a mass originating from Higgsing the $\Uoneprime$.
After that we compute the atomic ionization rates from dark photons,
taking full account of the medium effects, to derive powerful
constraints on the parameter space of the model using the results of
the XENON10 experiment.

\paragraph{Dark Photons}
\label{sec:HP}

The minimal extension of the SM gauge group by an additional
$\Uoneprime$ gauge factor yields the following effective Lagrangian
well below the electroweak scale,
\begin{align}
  \label{eq:L}
  \mathcal{L} = -\frac{1}{4} F_{\mu\nu}^2-\frac{1}{4} V_{\mu\nu}^2 -
  \frac{\kappa}{2} F_{\mu\nu}V^{\mu\nu} + \frac{m_V^2}{2} V_{\mu}V^{\mu}
  + e J_{\mathrm{em}}^{\mu} A_{\mu} ,
\end{align}
where $V_{\mu}$ is the vector field associated with the Abelian factor
$\Uoneprime$. The field strengths of the photon $F_{\mu\nu} $ and of
the dark photon $ V_{\mu\nu}$ are connected via the kinetic mixing
parameter $\kappa$ where a dependence on the weak mixing angle was
absorbed;  $ J_{\mathrm{em}}^{\mu}$ is the usual electromagnetic
current with electric charge~$e$.

Because of the U(1) nature of (\ref{eq:L}), we must distinguish two
cases for the origin of $m_V$: the Stueckelberg case (SC) with
non-dynamical mass, and the Higgs case (HC), where $m_V$ originates
through the spontaneous breaking of \Uoneprime. In the latter case,
(\ref{eq:L}) is extended by, $\mathcal{L}_{\phi} = |D_{\mu}\phi|^2 -
V(\phi)$ with the dark Higgs field $\phi = 1/\sqrt{2}(v'+ h')$ in
unitary gauge and after spontaneous symmetry breaking. The \Uoneprime\
covariant derivative is $D_{\mu}= \partial_{\mu}+i e' V_{\mu}$,
so that $m_V = e'v'$.  The interactions between the physical field
$h'$ and $V_{\mu}$ are given by,
\begin{align}
  {\cal L}_{\rm int}= e' m_V h' V_\mu^2
  +\frac{1}{2} e'^2\,h'^2 V_\mu^2 .
\end{align}
The crucial difference between the two cases comes in the small $m_V$
limit: while all processes of production or absorption of $V$ in SC
are suppressed, $\Gamma_{\rm SC} \sim O(m_V^2)$, in HC there is no
decoupling, and $\Gamma_{\rm HC} \sim O(m_V^0)$. Indeed, in the limit
$m_{V,h'}\to 0$ the $V$-$h'$ interaction with external electromagnetic
(EM) charge is equivalent to the interaction of charged scalar field
quanta with the effective EM charge of $e_{\rm eff} = \kappa e'$
\cite{Holdom:1985ag,Davidson:2000hf}. Thus, the emission of particles
from the U(1)$_V$ sector is generically given by
\begin{align} 
{\rm SC:} ~~ \gamma^{(*)} \to V ;~~{\rm HC:} ~~
  \gamma^{(*)} \to Vh',
\end{align}
where $\gamma^{(*)}$ is any---virtual or real---photon.  The
ionization of an atom $A$ in the detector can then be schematically
described as
\begin{align}
\label{absorb}
{\rm SC:} & ~~  V + A \to A^+ +e^- ,~~~~~~~~\\
{\rm HC:} & ~~  V(h') + A \to h'(V) +A^+ +e^-,
\label{absorbHC}
\end{align}
where again all interactions are mediated by $\gamma^{(*)}$.

\paragraph{Solar flux}
\label{sec:flux}

\begin{figure}[tb]
\centering
\includegraphics[width=\columnwidth]{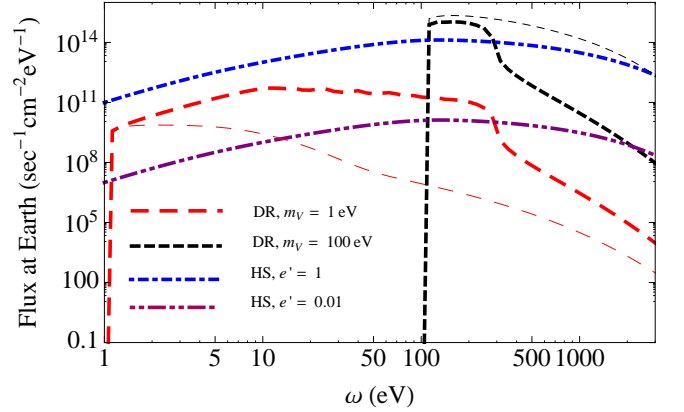}
\caption{Fluxes at the Earth as functions of energy for both
  the SC and HC dark photon for $\kappa = 10^{-12}$. The red and black thick dashed
  curves show the contribution from longitudinal dark radiation (DR) for $m_V=$1 eV
  and 100 eV, respectively. The corresponding thin curves show the transverse contribution. The blue and purple dotted dashed curves show the contribution from the Higgs-strahlung (HS) process for $e'=1$ and 0.01, respectively.}
\label{fig:flux}
\end{figure}

The solar flux of dark photons in the SC is thoroughly calculated in
Ref.~\cite{An:2013yfc}.  In the small mass region, $m_V\ll\omega_p$
where $\omega_p$ is the plasma frequency, the emission of longitudinal
modes of $V$ dominates the total flux, and the emission power of dark
photons per volume can be approximated as
\begin{equation}
\label{res}
\frac{d P_L}{d V } \approx \frac{1}{4\pi} \frac{\kappa^2 m_V^2 \omega_p^3}{e^{\omega_p/T} - 1}  .
\end{equation} 
For the purpose of this paper, a more useful quantity is the
energy-differential flux of dark photons at the location of the Earth.  The
spectra for some representative values of the parameters are shown in
Fig.~\ref{fig:flux}.

We now turn to the HC: as already mentioned, in the small $m_V$
region, the Higgs-strahlung process dominates the flux, whereas in the
region where $m_V$ is comparable to the plasma frequency inside the
Sun, $\omega_p = O(100\,\eV)$, this process is subdominant due to
phase space suppression.  In vacuum only an off-shell photon can
convert to $V$. Inside a medium, however, the pole position is shifted and
the $\gamma^{(*)}\to V$ process is equivalent to the decay of either a
``massive'' transverse mode or a (longitudinal) plasmon.  Inside the
Sun, since transverse photons are more numerous than plasmons
($\omega_p^3\ll T^3$), the Higgs-strahlung process is dominated by
the decay of transverse photons.  The corresponding matrix element can
be written as
\begin{equation}\label{element1}
{\cal M} = e'\kappa \epsilon^T_\mu(q) (k_1 - k_2)^\mu \ ,
\end{equation}
where $k_1$ and $k_2$ are the four-momenta of the outgoing dark Higgs,
$q = k_1 + k_2$ is the four-momentum of the decaying photon with
transverse polarization vector $ \epsilon^T_\mu$, and $q^2 \approx
\omega_p^2$. Therefore, the total energy power density of dark
radiation contributed by the Higgs-strahlung (HS) process can be estimated
as
\begin{equation}
  \left.\frac{dP}{dV}\right|_{HS} \approx \int\frac{ d\Phi_2 d^3 \vec q}{2q^0 (2\pi)^3} 
\frac{2 q^0}{e^\frac{q^0}{T} - 1} \overline{|{\cal M}|^2} = 
\frac{e_{\rm eff}^2 \omega_p^5}{48\pi^3} f \left(\frac{\omega_p}{T}\right) ,
\end{equation}
where $d\Phi_2$ is the two-body phase space of the final state,
$\overline{|{\cal M}|^2}$ averages over the polarization of the
transverse photons, and $f(a) = \int_1^{\infty} dx
(x^2-1)^{1/2}x/(e^{ax}-1)$. From the matrix element (\ref{element1}),
we can also calculate the joint differential production rate of the
dark vectors and Higgses, which can be written as
\begin{eqnarray}
  \left.\frac{d\Gamma^\phi}{dV d\omega}\right|_{HS} = \frac{e_{\rm eff}^2  \omega_p^2}{4\pi^3 } 
  \int_{\omega+\frac{\omega_p^2}{4\omega}}^\infty \frac{dq^0 (\omega q^0 - \omega^2 - {\omega_p^2}/{4})}{(e^{q^0/T} - 1)[{(q^0)}^2-\omega_p^2]} \ .
\end{eqnarray}
It is important to note that for small $m_{V}$ if medium effects
restore the U(1)$_V$ symmetry by driving $v'\to 0$, the
Higgs-strahlung rate remains valid.  The flux of dark photons on
the Earth in the HC for small $m_{V(h')}$ is also shown in
Fig.~\ref{fig:flux}.  As can be seen, the flux of $V\, (h')$ is not
enhanced in the IR but rather attains a broad maximum at dark photon energies
$\omega \sim$~100~eV.

\paragraph{Absorption of dark photons}
\label{sec:DDS}

To calculate the absorption rate of dark photons in the detector's
material (\ref{absorb}), we need to know the photo-electric absorption
cross section $\sigma_{abs}$ and the index of refraction, encoded in
the real and imaginary part of $\varepsilon_r$, the relative
permittivity of the target material.

In the SC, the amplitude for the absorption of a dark photon consists of
the atomic transition matrix element multiplied by the propagator of
$\gamma^{(*)}$. According to Ref.~\cite{An:2013yfc}, it can be written
as
\begin{equation}\label{amplitude1}
{\cal M}_{i\rightarrow f+V_{T,L}} = - \frac{\kappa m_V^2}{m_V^2 - \Pi_{T,L}} \langle f| [e J^\mu_{\rm em}]|i\rangle \epsilon^{T,L}_\mu \ ,
\end{equation}
where $\epsilon_{\mu}^{T,L}$ are the polarization vectors for the
transverse and longitudinal modes of the dark photon, ($\epsilon_\mu^2=-1$), and
$\Pi_{T,L}$ are defined via the polarization tensor inside the medium
of the detector:
\begin{equation}
\Pi^{\mu\nu} \equiv e^2 \langle J^{\mu\dagger}_{\rm em},J^\nu_{\rm em}\rangle
= \Pi_T \sum_{i=1,2}{\epsilon^T_i}^\mu {\epsilon^T_i}^\nu + \Pi_L \epsilon^{L\mu} \epsilon^{L\nu} \ .
\end{equation}
The total absorption rate can be written as
\begin{equation}
  \Gamma_{T,L}
  = \frac{\kappa_{T,L}^2 e^2 \epsilon^{T,L*}_\mu \epsilon^{T,L}_\nu}{2\omega} 
  \int d^4 x e^{iq\cdot x} \langle i| J_{\rm em}^{\mu\dagger}(x) J_{\rm em}^\nu(0) |i\rangle,
  \label{Gamma}
\end{equation}
where $q$ is the dark photon four-momentum with $\omega \equiv q^0$ and $\kappa_{T,L}$ are the effective mixings for the transverse and longitudinal modes respectively,
\begin{eqnarray}\label{kappa}
\kappa_{T,L}^{2} = \frac{\kappa^{2} m_{V}^{4}}{(m_{V}^{2} - \real \Pi_{T,L})^{2} + (\imag \Pi_{T,L})^{2}} \ . 
\end{eqnarray}
In~(\ref{Gamma}), the correlation function should be taken in the
physical region $\omega>0$ where it is equal to $- 2{\rm Im} \langle
J_{\rm em}^{\mu\dagger}, J_{\rm em}^\nu \rangle = e^{-2} {\rm Im}
\Pi^{\mu\nu}$ by unitarity (see, {\em e.g.}  \cite{IZ}).
Therefore, the total absorption rate can be simplified to
\begin{equation}\label{gamma}
\Gamma_{T,L} = - \frac{\kappa_{T,L}^2 {\rm Im \Pi_{T,L}}}{\omega} \ .
\end{equation}
Finally, in an isotropic non-magnetic material one has 
\begin{equation}
\Pi_T = - \omega^2 \Delta\varepsilon_r  \ , \;\; \Pi_L = - q^2 \Delta\varepsilon_r \ ,
\end{equation}
where $\Delta\varepsilon_r \equiv \varepsilon_r-1$.  Combining
Eqs.~(\ref{kappa}) and (\ref{gamma}) we build the main formulae for
the absorption rates of the transverse and longitudinal modes:
\begin{eqnarray}
\Gamma_T &=& { \left( \frac{\kappa^2m_V^4{\imag \varepsilon_r}}{\omega^3|\Delta\varepsilon_r|^2} \right)  }
{\left[1 + \frac{2 m_V^2 \omega^2 \real\Delta\varepsilon_r + m_V^4}{\omega^4 |\Delta\varepsilon_r|^2}\right]^{-1}} \!\!\!\!\! , 
\nonumber\\
\Gamma_L &=& \frac{\kappa^2 m_V^2 \imag \varepsilon_r}{\omega|\varepsilon_r|^2} \ .
\label{gamma2}\end{eqnarray}
In general, $\varepsilon_r$ depends on both the
injecting energy $\omega$ and $\vec q^2$.  The latter is suppressed by
$\sim \vec q^2/(\omega m_e)$ and to good accuracy can be neglected.
One can see that in the region $m_V^2 \ll \omega^2
|\Delta\varepsilon_r|$, the absorption rate of the $T$-modes scales as
$m_V^4$ whereas for the $L$-mode, it is always proportional to
$m_V^2$. In the opposite limit, Eq.~(\ref{gamma2}) is given by the
number density of atoms $n_A$, $\sigma_{abs}$, and the velocity of
dark photons $v_V$, $\Gamma_T = \kappa^2 \omega \imag \varepsilon_r
= \kappa^2 n_A v_V^{-1}\sigma_{abs} $, (we work in $c=1$ units).

Going over to the HC, we take $m_{h'}\sim m_V$, and consider the
absorption process (\ref{absorbHC}) in the limit of both masses being
small.  Using the equivalence to the scattering of charged scalars, we
write the amplitude as
\begin{equation}
{\cal M} = e_{\rm eff}
 (k_1 + k_2)^\mu \langle A_\mu,A_\nu\rangle \langle f | J^\nu_{\rm em}(q) |i\rangle \ ,
\end{equation}
where $k_1$ and $k_2$ are the four momenta of the incoming and
outgoing $\phi$ particles and $q = k_1 - k_2$.  Medium-corrected
photon propagators $\langle A_\mu,A_\nu\rangle$ in the Coulomb gauge
are given by
\begin{eqnarray}
\langle A^i, A^j \rangle = \frac{\delta^{ij} - \hat q^i \hat q^j}{q^2 - \Pi_T}  \ , \;\langle A^0, A^0 \rangle = \frac{q^2}{|\vec q|^2\left(q^2 - \Pi_L \right)} \ ,
\end{eqnarray}
where $\hat q\equiv \vec q/|\vec q|$. Following the same steps as in
the SC, summing over all the possible excited atomic states in the
medium, we get
\begin{equation}
\sum_f |{\cal M}|^2 \approx - 8  e_{\rm eff}^2   { \imag \varepsilon_r} 
\frac{q^2 k_1^0( k_1^0 - q^0 )}{|q^2 + {q^0}^2 \Delta\varepsilon_r|^2} \ ,
\end{equation}
where terms with further suppression by $\Delta \varepsilon_r$ are
omitted. The differential scattering rate with respect to the energy
transfer to atoms $q^0$ is given by,
\begin{equation}
\label{GHC}
\frac{d\Gamma}{d q^0} \approx \frac{ e_{\rm eff}^2 }{4\pi^2} \frac{k_1^0 - q^0}{k_1^0} 
\left[ \log\left( \frac{4k_1^0(k_1^0 - q^0)}{{(q^0)}^2 |\Delta\varepsilon_r|} \right) - 1\right] \imag\varepsilon_r(q^0) ,
\end{equation}
Notice that the collinear divergence is regularized by the in-medium
modification of the photon propagator.

\paragraph{Limits from direct detection}
\label{sec:DDS}

Having obtained both solar fluxes and the absorption rates of dark
photons, we are ready to calculate the experimental event rate. In a
given experiment, the expected number of signal events in the SC can
be written as
\begin{equation}\label{master1}
N_{\rm exp} = V T \int^{\omega_{\rm max}}_{\omega_{\rm min}} \frac{\omega d \omega}{|\vec{q}|} \left(\frac{d\Phi_{T}}{d\omega} \Gamma_T + \frac{d\Phi_{L}}{d\omega} \Gamma_L\right) {\rm Br}  ,
\end{equation}
where $V$ and $T$ are the fiducial volume and live time of the
experiment, respectively, and ${\rm Br}$ is the branching ratio of
photoionization rate to total absorption rate.

Since both $\real\varepsilon_r$ and $\imag\varepsilon_r$ are
proportional to the number density of atoms of the material, $n_A$, in
the small $m_V$ limit $\Gamma_T\propto n_A^{-1}$.  As a result, low
density materials are best suited for the detection of $T$-modes. 
However, as discussed in Ref.~\cite{An:2013yfc}, the major
component of the dark photon flux from the Sun is longitudinal, and
from Eq.~(\ref{gamma2}) we have $\Gamma_L\propto n_A$. Therefore, the
detection abilities are directly proportional to the total active mass
inside the detector.  Given the significant enhancement in the
low-energy part of the solar dark photon spectrum,
Fig.~\ref{fig:flux}, a detector with a low threshold energy of $O(100)
$~eV will have a clear advantage.  
To date, the only work that considers limits on dark photons from
direct DM detection is by HPGe collaboration,
Ref.~\cite{Horvat:2012yv}. 
However, it used incomplete calculations of the solar flux, and as we
will show in the following, the low-energy ionization signals by the
XENON10 ~\cite{Angle:2011th} and CoGeNT ~\cite{Aalseth:2011wp,Aalseth:2012if}
collaborations yield far more stringent limits.

\begin{figure}[tb]
\centering
\includegraphics[width=\columnwidth]{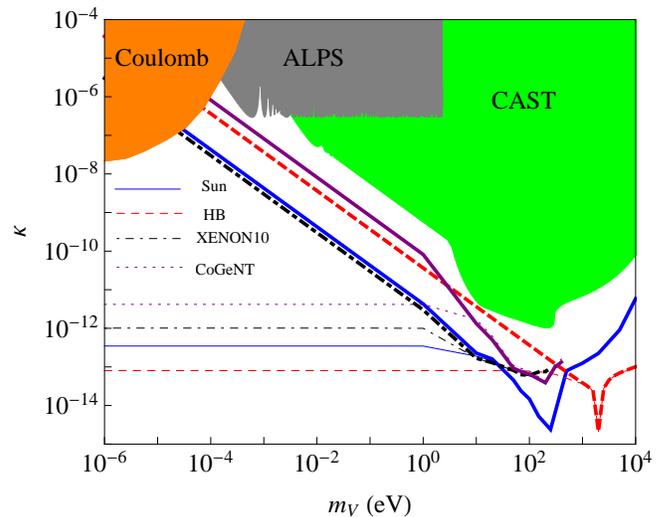}
\caption{Constraints on $\kappa$ as functions of $m_V$. The solid,
  dashed, dot-dashed and dotted curves show constraints from the
  energy loss of the Sun by requiring that the dark photon luminosty
  does not exceed 10\% of the standard solar
  luminosity~\cite{Gondolo:2008dd}, energy loss in horizontal branch (HB) stars, the XENON10 experiment and the CoGeNT experiment,
  respectively. The thick curves are for the SC, whereas the thin
  curves are for the HC with $e'=0.1$. For comparison, the current
  bound (gray shading) from the LSW-type experiments are shown (see
  Ref. \cite{Ehret:2010mh} for details). The conservative constraint
  from the CAST experiment~\cite{Andriamonje:2007ew} by considering
  the contributions from only the transverse
  modes~\cite{Redondo:2008aa} is also shown in green shading.The
  orange shaded region is excluded from tests of the inverse square
  law of the Coulomb interaction~\cite{Bartlett:1988yy}.}
\label{fig:plotall}
\end{figure}

The XENON10 collaboration has published a study on low-energy
ionization events in~\cite{Angle:2011th}. With 12.1~eV ionization
energy, the absorption of a dark photon with 300~eV energy can produce
about 25 electrons. To get a conservative constraint we count all the
ionization events within 20~keV nuclear recoil equivalent in
Ref.~\cite{Angle:2011th}, which corresponds to a signal of about 80 electrons. The
total number of events is 246, which indicates a 90\% C.L upper limit
on the detecting rate to be $r < 19.3$ events kg$^{-1}$day$^{-1}$ (similar to
limits deduced in Ref. \cite{Essig:2012yx}).  In the region $12.1{~\rm
  eV}<\omega<300$~eV the ionization process dominates the absorption,
and therefore ${\rm Br}$ in this region can be set to unity. The 90\%
C.L. upper limit on $\kappa$ as a function of $m_V$ is shown by the
dot-dashed black curve in Fig.~\ref{fig:plotall}, where we can see
that it gives the most stringent constraint in the SC. To arrive at
these limits we reconstruct $\varepsilon_r$ for Xe from the real and imaginary parts of the refractive index of Xe. The imaginary part can be extracted from the total photoabsorption cross section~\cite{Henke,oscillationlength}, and the real part can be calculated from the imaginary part using the Kramers-Kronig dispersion relations. 
The improvement over other experimental
probes is quite significant, considering that the signal scales as
$\kappa^4$. We also collate main constraints in Table~\ref{table:exps}.

The published data from the CoGeNT DM experiment have a threshold of
about 450~eV~\cite{Aalseth:2012if}. In this region, the dark photon
flux from the Sun drops almost exponentially with energy, whereas the
observed spectrum in CoGeNT is relatively flat. Therefore, in order to
optimize the sensitivity, we only use the event counts in the interval
$450-500$~eV and the 90\% upper limit on the background subtracted
rate is $r < 0.6$ events kg$^{-1}$day$^{-1}$. The resulting
sensitivity is shown as the thick dotted purple curve in
Fig.~\ref{fig:plotall}, which is far weaker than the constraint from
the energy loss of the Sun. 

In the HC, $N_{\rm exp}$ in Eq.~(\ref{master1}) must include a
contribution from (\ref{GHC}), and it dominates in the region $m_V^2
\ll \omega^2 |\Delta\varepsilon_r|$, but is subdominant if $m_V^2\sim
\omega^2 |\Delta\varepsilon_r|$.  Since the flux of $V(h')$ in the HC
is mainly contributed from conversion of transverse photons in the
Sun, the spectral distribution reflects the solar temperature, Fig.~\ref{fig:flux},
with the cutoff above 1~keV.  A dark photon of 1~keV energy can
at most produce about 60 electrons in liquid xenon.
For $e'=0.1$, the 90\% C.L. upper limit is shown as the thin
dot-dashed curve in Fig.~\ref{fig:plotall}. For the sensitivity from
CoGeNT we take into account all the events from 450~eV to 1~keV and
the associated line is shown as the thin dotted curve in
Fig.~\ref{fig:plotall}, and included in Table~\ref{table:exps} as limit on $e_{\rm eff}$.
In both, the HC and the SC, CoGeNT does not have sensitivity to constrain
dark photons since the required flux is not supported by the Sun.

\begin{table}[tb]
\caption{Sensitivities to $\kappa$ and $e_{\rm eff}$ in the small $m_V$ region.}
\begin{center}
\begin{tabular}{lcccc}
\hline
Model param. & Sun & HB & XENON10 & CoGeNT \\
\hline
\hline 
SC, $\kappa \times \frac{m_V}{\rm eV} $ &   $4\times10^{-12}$&$4\times10^{-11}$&$3\times10^{-12}$&$8\times10^{-11}$\\
HC,  $e_{\rm eff}$ &  $3\times10^{-14}$&$8\times10^{-15}$&$1\times10^{-13}$&$4\times10^{-13}$   \\ 
\hline
\end{tabular}
\end{center}
\label{table:exps}
\end{table}%

\paragraph{Conclusions}
\label{sec:conclusions}

We point out that the unprecedented sensitivity of some of the DM
experiments to ionization allows to turn them into the most sensitive
dark photon helioscopes. By directly calculating the ionization
signal, we show that the ensuing constraint from the XENON10
experiment significantly surpasses any other bounds on dark photons,
including very tight stellar energy loss constraints in the $m_V$-interval
from $10^{-5}$ to $100$~eV. In the case of ``mini-charged'' particles
(equivalent to the Higgsed version of dark photons), we also derive a
stringent bound, $e_{\rm eff} < 10^{-13}$, which is second only to the constraint from the energy loss of the horizontal branch stars; see also~\cite{Ahlers:2008qc}.  Given
the enormous amount of experimental progress in the field of direct DM
detection, one can be optimistic that future sensitivity to dark
photons, and other light particles is bound to be further improved.

\paragraph{Acknowledgements} We acknowledge useful correspondence with
G.~Raffelt and J.~Redondo on stellar production rates and the strength of the solar luminosity constraint. Research at the Perimeter Institute is
supported in part by the Government of Canada through NSERC and by the
Province of Ontario through MEDT.

\end{document}